\documentclass[sigplan,10pt,screen]{acmart}
\usepackage{xspace}
\usepackage{tabularx}
\usepackage{subfig}
\usepackage{enumitem}
\usepackage{algorithm}
\usepackage{algorithmicx}
\usepackage[noend]{algpseudocode}
\usepackage{enumitem}
\usepackage{pifont}
\setitemize{noitemsep,topsep=0pt,parsep=0pt,partopsep=0pt}

\newcommand{\parabf}[1]{\medskip\noindent\textbf{#1}}

\newcommand{\paraf}[1]{\noindent\textbf{#1}}
\newcommand{\cut}[1]{}

\newcommand{\sysname}{LoongServe\xspace}

\begin{document}
\title{\sysname: Efficiently Serving Long-Context Large Language Models with Elastic Sequence Parallelism}
\pagestyle{empty}

\author{Bingyang Wu}
\affiliation{%
  \institution{School of Computer Science\\Peking University}
  \country{}
}

\author{Shengyu Liu}
\affiliation{%
  \institution{School of Computer Science\\Peking University}
  \country{}
}

\author{Yinmin Zhong}
\affiliation{%
  \institution{School of Computer Science\\Peking University}
  \country{}
}

\author{Peng Sun}
\affiliation{%
  \institution{Shanghai AI Lab}
  \country{}
}

\author{Xuanzhe Liu}
\affiliation{%
  \institution{School of Computer Science\\Peking University}
  \country{}
}

\author{Xin Jin}
\affiliation{%
  \institution{School of Computer Science\\Peking University}
  \country{}
}

\begin{abstract}

The context window of large language models (LLMs) is rapidly increasing, leading to a huge variance in resource usage between different requests as well as between different phases of the same request. Restricted by static parallelism strategies, existing LLM serving systems cannot efficiently utilize the underlying resources to serve variable-length requests in different phases. To address this problem, we propose a new parallelism paradigm, \textit{elastic sequence parallelism} (ESP), to elastically adapt to the variance across different requests and phases. Based on ESP, we design and build \sysname, an LLM serving system that (1) improves computation efficiency by elastically adjusting the degree of parallelism in real-time, (2) improves communication efficiency by reducing key-value cache migration overhead and overlapping partial decoding communication with computation, and (3) improves GPU memory efficiency by reducing key-value cache fragmentation across instances. Our evaluation under diverse real-world datasets shows that \sysname improves the throughput by up to 3.85$\times$ compared to chunked prefill and 5.81$\times$ compared to prefill-decoding disaggregation.
\end{abstract}



\acmYear{2024}\copyrightyear{2024}
\setcopyright{acmlicensed}
\acmConference[SOSP '24]{ACM SIGOPS 30th Symposium on Operating Systems Principles}{November 4--6, 2024}{Austin, TX, USA}
\acmBooktitle{ACM SIGOPS 30th Symposium on Operating Systems Principles (SOSP '24), November 4--6, 2024, Austin, TX, USA}
\acmDOI{10.1145/3694715.3695948}
\acmISBN{979-8-4007-1251-7/24/11}

\maketitle

\thispagestyle{empty}

\section{Introduction}
\label{sec:introduction}

The emergence of large language models (LLMs) has fundamentally pushed modern applications into a new era. These LLMs are trained on vast data to gain extraordinary capabilities in a wide range of areas, such as programming~\cite{nijkamp2023codegen}, chatting~\cite{chatgpt} and planning~\cite{significantgravitasautoGPT}.
The context window is a key feature of LLMs.
It is rapidly increasing to 
enable sophisticated reasoning about long documentation, relevant problem-solving with a large codebase, and customized generation based on long instructions~\cite{gemini}. Many organizations have released long-context LLMs, such as Anthropic's Claude-3~\cite{claude3}, Google's Gemini-1.5~\cite{gemini}, and UC Berkeley's Large World Model (LWM)~\cite{liu2023world}, which all support a 1M context window.

Serving long-context LLMs poses significant challenges to LLM serving systems in both GPU computing and GPU memory. During the serving process, the GPU memory consumption of key-value caches grows linearly with the sequence length. When serving a \textit{single} request with the input length of 1M tokens by a LWM model, the key-value cache alone can amount to 488GB, far exceeding the GPU memory capacity of the most advanced GPUs available today. As for the computational demand, the complexity of the attention mechanism in advanced long-context LLMs, such as LWM, is quadratic to the input sequence length, making the processing of long sequences more computationally intensive.

To accelerate the LLM computation, many works exploit different parallel
dimensions. \textit{Model parallelism}~\cite{shoeybi2020megatronlm} partitions
the model parameters across multiple GPUs to parallelize the computation.
\textit{Sequence parallelism}~\cite{korthikanti2022reducing} partitions input
sequences of requests across GPUs to achieve acceleration.
Whether using one of them or a combination of both, existing practices decide the parallel configuration \emph{statically} before launching the service.

However, the LLM inference workloads are highly dynamic. As the context window of LLMs increases, the variance of input lengths of requests becomes larger, leading to distinct computational demands for different requests. Furthermore, request processing is divided into two phases, the \textit{prefill} phase and the \textit{decoding} phase. The resource demand of the same request in different phases also varies significantly~\cite{agrawal2023sarathi, holmes2024deepspeedfastgen, patel2023splitwise, zhong2024distserve, hu2024inference,wu2023fast}. Therefore, existing static parallelisms are not efficient for requests with varied input lengths, nor for the different phases of the same request.

One possible solution to mitigate these problems is to organize GPUs into multiple groups. Each group deploys an LLM instance and adopts a different parallel strategy to process sequences in a specific range of sequence length or a specific phase~\cite{zhong2024distserve, hu2024inference, patel2023splitwise}.

However, their static grouping strategy often mismatches with the resource demand of varied requests in different phases, because the resource demand dynamically changes at the granularity of iterations~\cite{yu2022orca}. Additionally, these solutions~\cite{zhong2024distserve, hu2024inference, patel2023splitwise} migrate key-value caches of \textit{all} requests when transiting the phase, incurring extra communication overhead. Due to isolation among groups, the GPU memory across different groups cannot be utilized together to serve requests with long sequence lengths, leading to GPU memory fragmentation (\S\ref{subsec:motivations}).

To fundamentally address these problems, we propose \textit{Elastic Sequence Parallelism} (ESP). Unlike existing static parallelisms, ESP dynamically decides the \textit{Degree of Parallelism} (DoP) for requests in each iteration. For instance, during the prefill phase, ESP could set the DoP to the total number of GPUs, leveraging the entire cluster's resources to quickly process the request. Upon transiting to the relatively lightweight decoding phase, ESP can decrease the DoP to reduce communication overhead and release unnecessary resources to accelerate the processing of other requests.

However, fully unleashing the potential of ESP is challenging. First, if the elastic scaling overhead is excessive, it can negate the benefits of flexible resource allocation that ESP provides. 
For instance, if ESP involves the migration of a substantial amount of key-value caches across GPUs when scaling, it will incur significant communication overhead. Second, the dynamic loads of requests with variable sequence lengths in different phases form a complicated scheduling space, while decisions need to be made at the granularity of iterations, whose duration can be the scale of tens of milliseconds. How to find an efficient scheduling plan with such strict latency requirements remains a challenging problem. 

To this end, we propose \sysname, the first LLM inference serving system
equipped with ESP to serve long-context LLMs. To realize efficient ESP, \sysname
adopts a set of novel elastic scaling mechanisms with no extra communication in
both scale-up and scale-down scenarios. For the prefill phase, we propose a
\emph{proactive scaling-down} mechanism that combines the prefill phase and
scaling down to reuse the communication of the prefill phase, thereby
eliminating extra communication overhead. For the decoding phase, we propose a
\emph{multi-master decoding} mechanism that avoids migrating existing key-value caches
and overlapping partial decoding communication with computation, thereby
reducing the communication overhead. All these mechanisms manage tokens at the
granularity of a single token across instances without any locality constraints,
thereby eliminating GPU memory fragmentation. For online scheduling, \sysname
uses a scalable four-step scheduling algorithm that considers
DoP setting, batching, key-value cache placement, and elastic scaling,
to make decisions at the granularity of iterations with a polynomial
complexity.

Experiments on real-world datasets show that \sysname improves the throughput by up to 3.85$\times$ compared to chunked prefill and 5.81$\times$ compared to prefill-decoding disaggregation. Furthermore, experiments show that \sysname improves the performance of the prefill phase and decoding phase simultaneously.

In summary, we make the following contributions:
\begin{itemize}[leftmargin=*]
    \item We identify the limitations of existing solutions in serving long-context LLMs and propose Elastic Sequence Parallelism (ESP) as a solution.
    \item We propose a set of efficient elastic scaling mechanisms and a scalable scheduling algorithm to unleash the potential of ESP.
    \item We evaluate \sysname comprehensively to show its effectiveness compared to state-of-the-art solutions.
\end{itemize}

\section{Background and Motivation}
\label{sec:background}

\begin{figure}[t]
    \includegraphics[width=\linewidth]{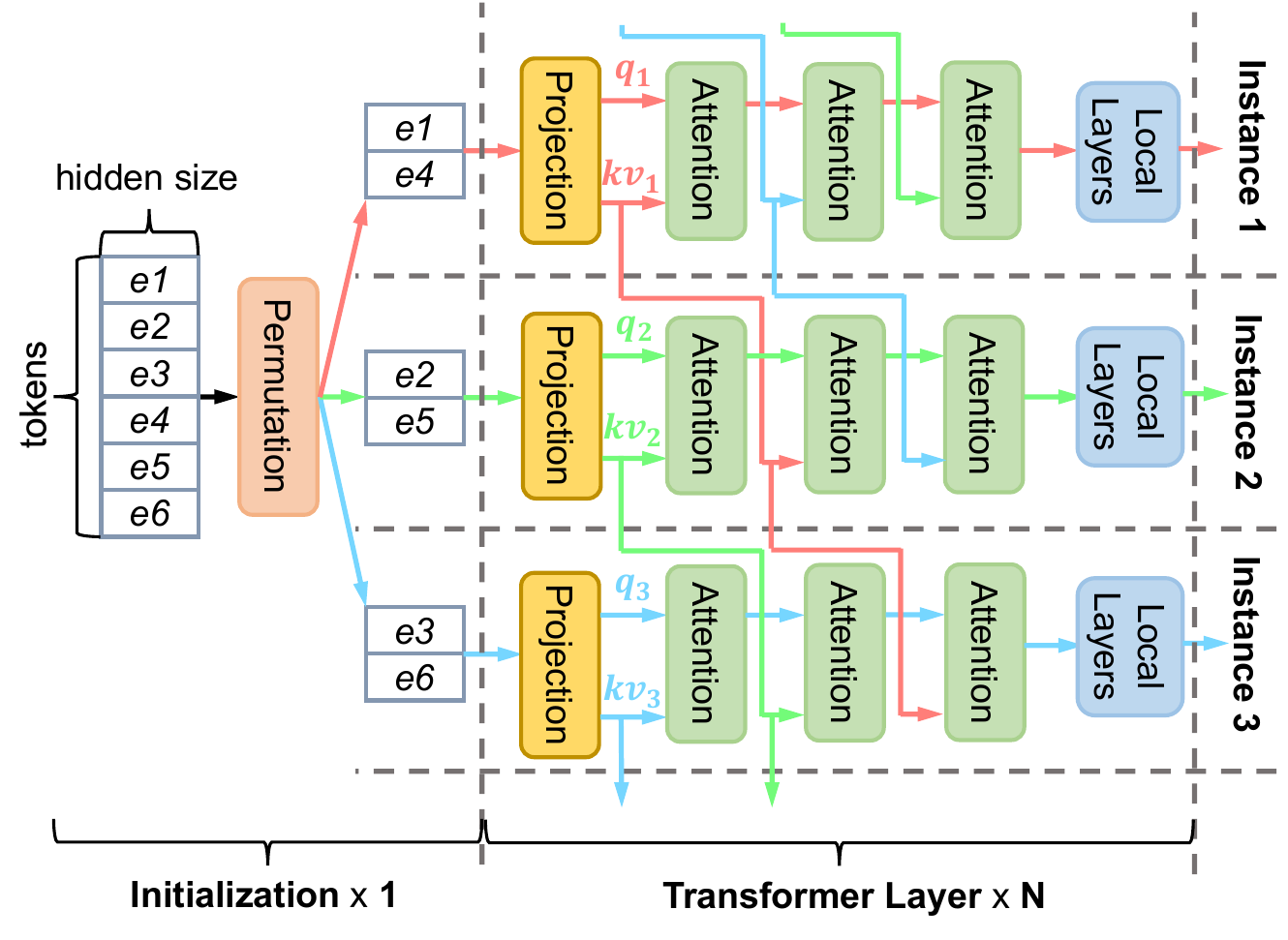}
    \vspace{-0.3in}
    \caption{Sequence parallelism in the prefill phase.}
    \label{fig:background:sequence_parallelism}
    \vspace{-0.1in}
\end{figure}

\subsection{The Process of LLM Inference}

Most of the popular LLMs~\cite{openai2023gpt4,claude3,bard,touvron2023llama,touvron2023llama2} adopt the Transformer architecture~\cite{vaswani2017attention}. A model typically consists of a stack of transformer layers, each containing an attention layer and a feed-forward network (FFN) layer. Attention layers make tokens in a request interact with other tokens, and FFN layers process requests token-wisely. In each iteration, given the tokens before it, the model predicts the next token. To avoid redundant computation, LLM serving systems cache the intermediate states of tokens, a.k.a \textit{key-value cache}, and reuse them for future token generation. This optimization divides the whole generation process into two phases: the \textit{prefill} phase and the \textit{decoding} phase. The prefill phase processes all the input tokens in a single iteration to build the key-value cache and generates the first output token, while the decoding phase only needs to compute the key-value cache for the newly generated output token. As a result, the prefill phase is more compute-intensive than the decoding phase.

\subsection{Existing LLM Serving Systems}
\label{subsec:existing_systems}

To accelerate LLM inference, existing solutions employ efficient GPU kernel implementations, e.g., Flash Attention~\cite{dao2022flashattention} and Flash Decoding~\cite{dao2023flashattention2} within a GPU, and exploit model parallelism, such as tensor parallelism~\cite{shoeybi2020megatronlm}, to partition model parameters across GPUs to parallelize the computation. However, the degree of model parallelism needs to be determined before launching the system.

To mitigate the impact of the long context, chunked prefill~\cite{patel2023splitwise, holmes2024deepspeedfastgen} splits the long context into chunks and processes them chunk by chunk with the decoding phase, but still incurs interference between two phases~\cite{zhong2024distserve}. To avoid interference, prefill-decoding disaggregation~\cite{zhong2024distserve} disaggregates two phases into different groups of GPUs. However, as mentioned in \S\ref{sec:introduction}, it leads to high communication overhead due to frequent migration, GPU computation inefficiency due to resource mismatch, and GPU memory fragmentation due to isolation between groups.

\begin{figure}[t]
    \includegraphics[width=\linewidth]{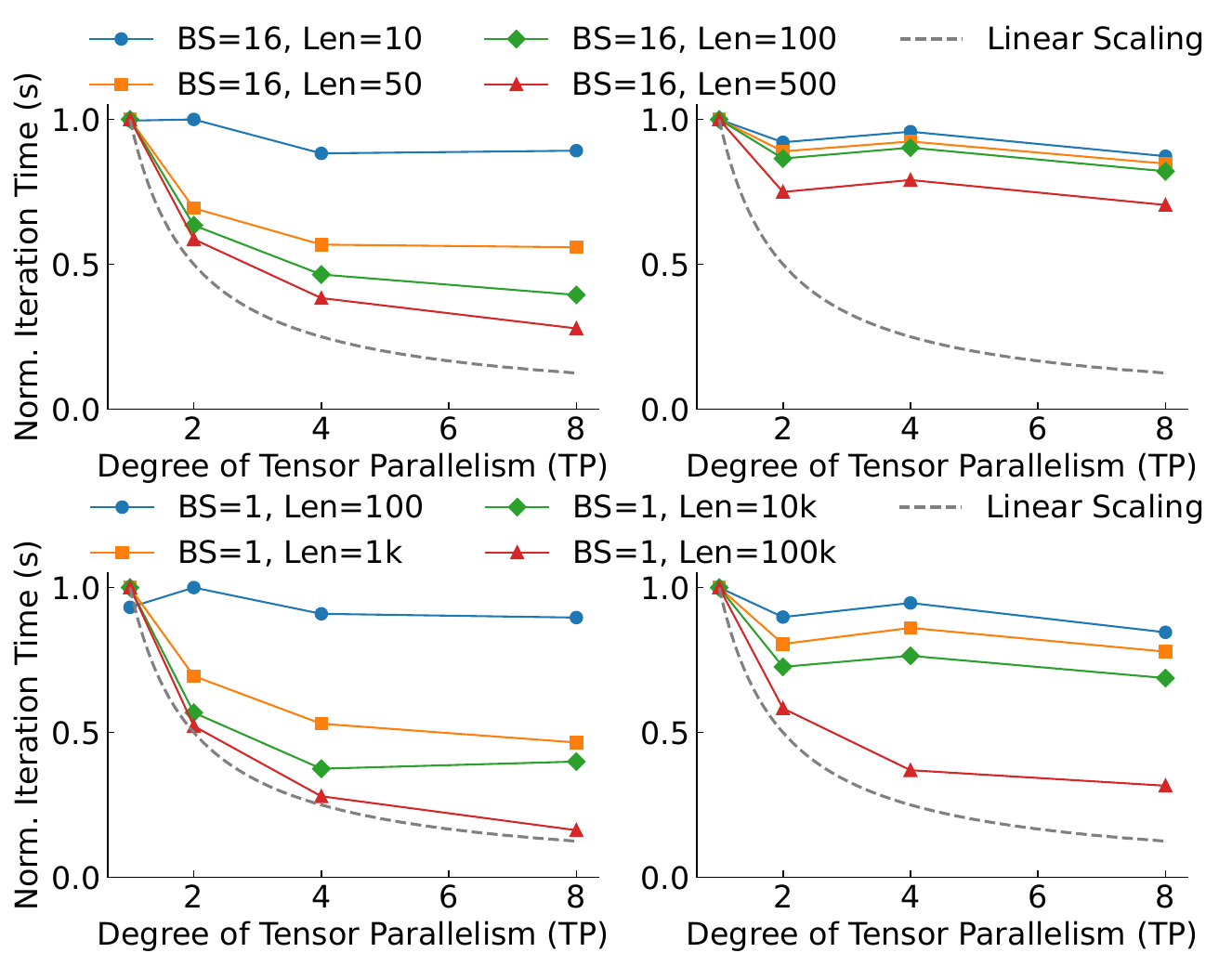}
    \vspace{-0.35in}
    \caption{Scalability of requests with different lengths in the different phases.}
    \label{fig:motivation:scaling_curves}
    \vspace{-0.1in}
\end{figure}

\subsection{Sequence Parallelism}
\label{sec:background:sequence_parallelism}
To accelerate long-context LLM training, many LLM training systems propose sequence parallelism~\cite{liu2023ring,brandon2023striped, ulysses, li2023lightseq, korthikanti2023reducing, li2023sequence}. In this work, we extend Striped Attention~\cite{brandon2023striped} to the LLM serving scenario.

\begin{figure}[t]
    \includegraphics[width=\linewidth]{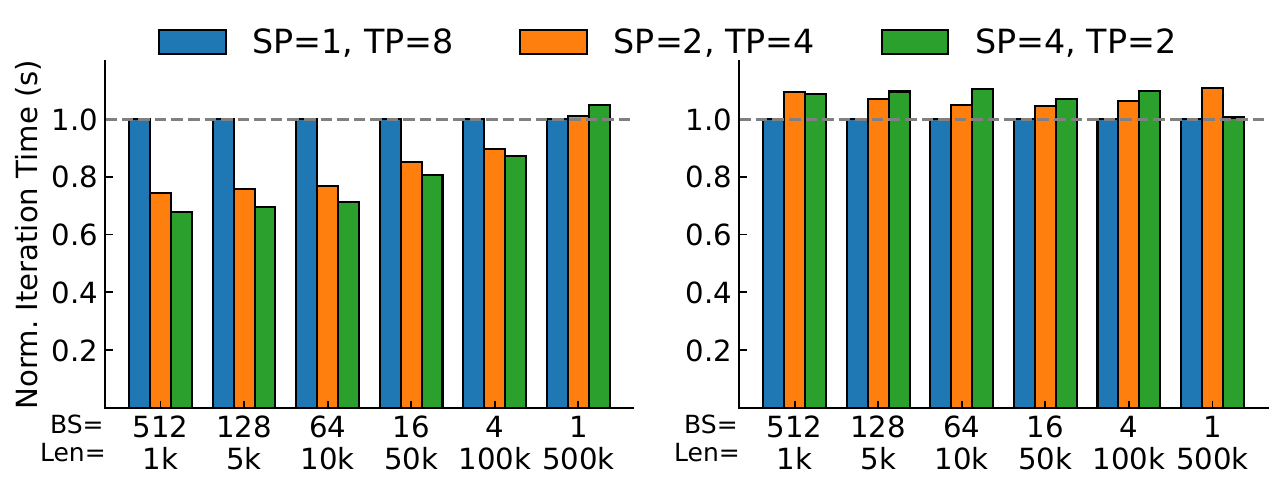}
    \vspace{-0.35in}
    \caption{Comparison between fixed sequence parallelism and tensor parallelism.}
    \label{fig:motivation:sp_performance}
    \vspace{-0.1in}
\end{figure}

Figure~\ref{fig:background:sequence_parallelism} illustrates its basic idea. Before processing, the input sequence is permuted, divided into multiple segments, and then dispatched to different instances. Except for the attention layer, the instances execute without communication. At the attention layer, each instance simultaneously (1) computes the attention with its local query ($q_i$) and key-value ($kv_j$) tensors and (2) sends the key-value tensors ($kv_j$) to its neighbor instance ($instance_{(i+1)\%n}$). After multiple rounds of this process, query tensors in each instance interact with all key-value tensors and then generate the final output to the next layer. Sequence parallelism is compatible with popular attention mechanisms, such as Multi-Head Attention (MHA)~\cite{vaswani2017attention}, Multi-Query Attention (MQA)~\cite{shazeer2019fast}, and Grouped-Query Attention (GQA)~\cite{ainslie2023gqa}. It also has the same computational complexity as tensor parallelism and consumes less GPU memory for buffer activations~\cite{liu2023ring}. Furthermore, it can be used in conjunction with other parallelisms, e.g., tensor parallelism, to accelerate LLM training further.

However, it cannot be directly applied to the LLM serving scenario. First, it only supports the prefill phase. Second, it is used in the LLM training scenario where the degree of parallelism in each training stage is fixed. Conversely, the LLM serving systems need to handle highly dynamic inference traffic and support unique characteristics introduced by the decoding phase, including unique computational patterns of the decoding phase and key-value cache management.

\subsection{Motivation and Challenges}
\label{subsec:motivations}

Despite existing LLM serving systems supporting long-context LLMs (\S\ref{subsec:existing_systems}), there is an inherent mismatch between their static nature and the dynamics in LLM serving. Particularly, LLM serving workloads are dynamic in two aspects. First, as the context window of LLMs increases, the resource demand during the prefill phase varies significantly across requests with different input lengths in both computation and GPU memory consumption. As shown in Figure~\ref{fig:motivation:scaling_curves}, processing 100K input tokens for a LWM model on 8 GPUs is 105.97 times slower than processing 1K input tokens. The difference in GPU memory consumption across requests can be as high as 1,000,000 times when serving an LLM with a 1M context window because the size of the key-value cache is linear to the input length. Secondly, even for a single request, the resource demand of the prefill and decoding phase differs vastly. As shown in Figure~\ref{fig:motivation:scaling_curves}, because the prefill phase is compute-intensive, increasing DoP can significantly accelerate it. Conversely, for the relatively compute-lightweight decoding phase, a larger DoP may lead to negligible performance improvement due to additional communication overhead. Therefore, static parallelism strategies cannot be efficient in all scenarios.

However, dynamically altering the parallelism strategy of LLM parameters, e.g., tensor parallelism, requires restarting the entire inference runtime, a process typically taking minutes and possibly longer for larger models, which is untenable for latency-sensitive serving scenarios. 

\begin{figure}[t]
    \includegraphics[width=0.85\linewidth]{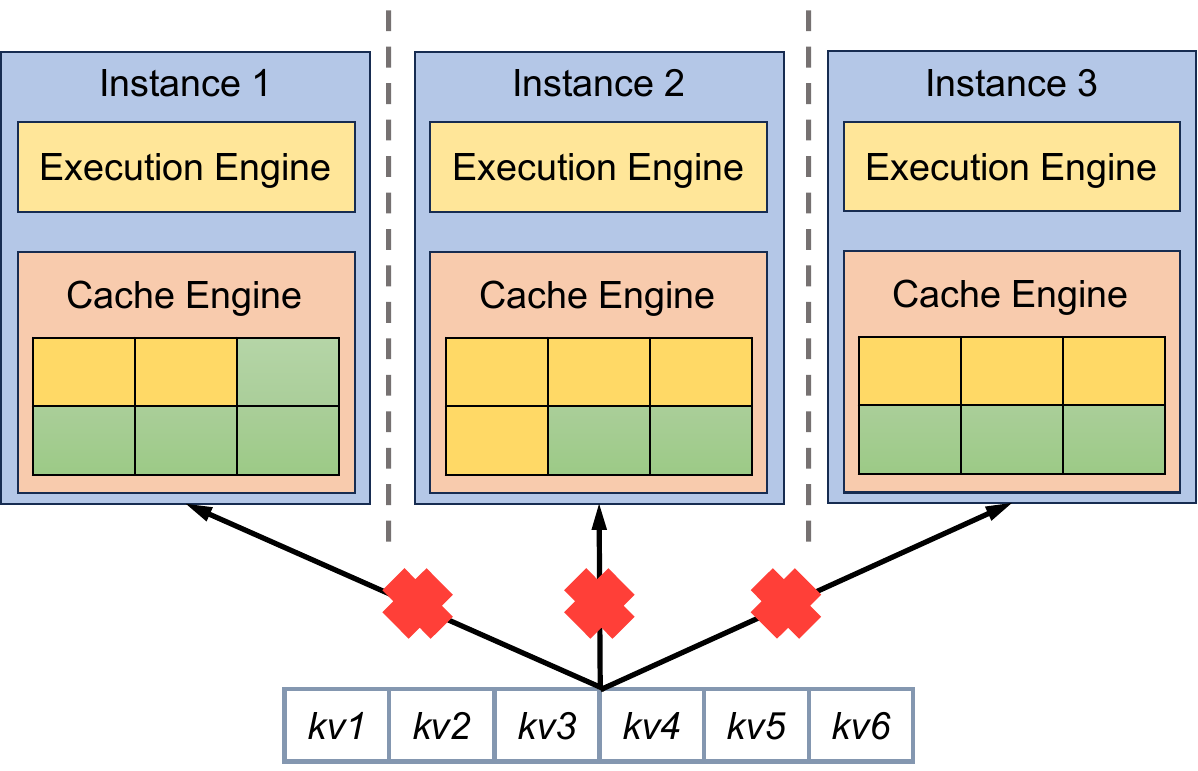}
    \vspace{-0.1in}
    \caption{KV cache fragmentation of group based strategy.}
    \label{fig:motivation:kv_cache_fragmentation}
    \vspace{-0.2in}
\end{figure}

Besides the inability to adapt to the dynamics, existing solutions~\cite{agrawal2023sarathi, holmes2024deepspeedfastgen, patel2023splitwise, zhong2024distserve, hu2024inference} also cause GPU fragmentation issues. The reason is that the entire or most of the key-value cache of a request must reside in a single instance due to the locality constraint. This constraint leads to situations as depicted in Figure~\ref{fig:motivation:kv_cache_fragmentation}. Despite there being sufficient overall GPU memory (six slots), none of the instances can serve a request with six tokens due to the locality constraint.

\parabf{Elastic sequence parallelism.}
To fundamentally address these issues, we propose elastic sequence parallelism (ESP). Our key insight is that LLM serving systems can extend SP to ESP by adding support to the decoding phase, flexibly managing the KV cache across phases, and dynamically distributing a request's input tokens across instances, allowing for an adjustable DoP without re-partitioning LLM parameters. As shown in Figure~\ref{fig:motivation:sp_performance}, using SP with existing parallelism strategies, such as tensor parallelism, does not introduce additional overheads and may even achieve better performance for requests with different lengths in both context and decoding phases, indicating a potential to further LLM inference acceleration. Moreover, ESP facilitates flexible utilization of memory resources across multiple instances, thereby mitigating the problem of memory fragmentation.

\parabf{Challenges.}
While the dynamic adjustment capability of ESP to meet the computational demands of diverse requests is indeed promising, unleashing its full potential presents several challenges.

\emph{Elasticity overhead.} In the LLM inference scenario, altering the degree of sequence parallelism implies the redistribution of key-value caches among instances. When serving requests with long contexts, the size of the key-value cache can be substantial. If each adjustment in SP incurs significant communication costs, such as frequent key-value cache migration, the advantages of ESP could easily be overshadowed. Hence, there is a critical need for an efficient and cost-effective migration mechanism.

\emph{Scheduling complexity.} In serving scenarios, the serving system faces hundreds of requests. It must decide on the grouping strategy, batching strategy, DoP of each batch, and request dispatching strategy for all requests in the system. Because the optimal scheduling is affected by the dynamic LLM inference workload, the scheduling decision must be made in this vast scheduling space within a latency constraint in the orders of tens of milliseconds.

\section{\sysname Overview}

\begin{figure}[t]
    \includegraphics[width=\linewidth]{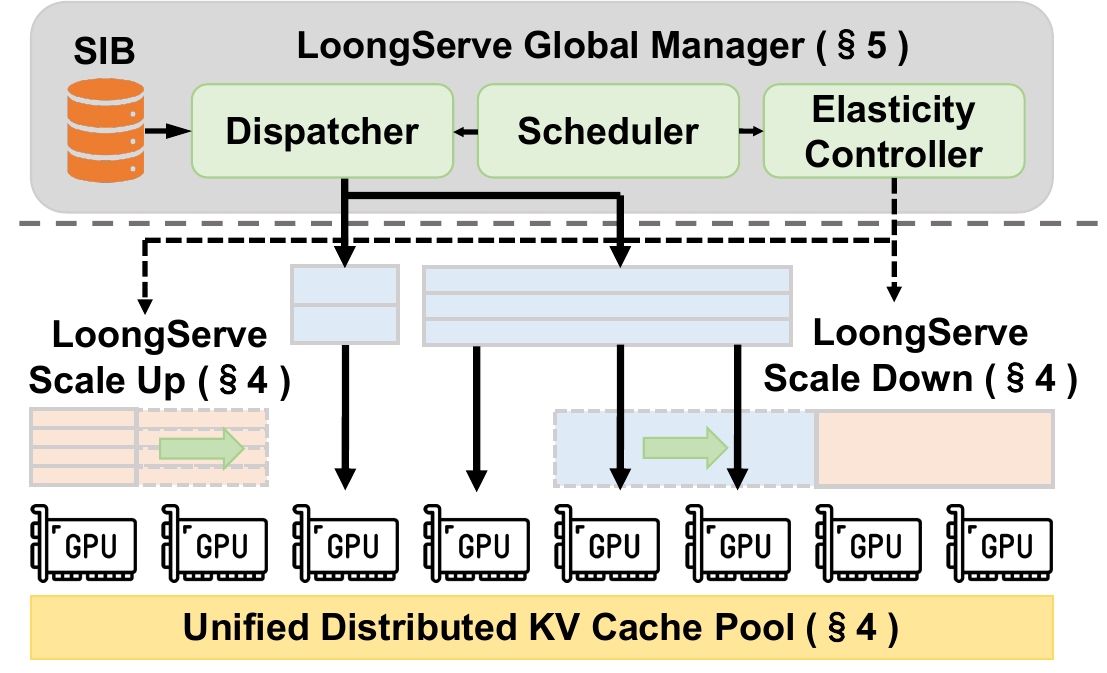}
    \vspace{-0.3in}
    \caption{\sysname system overview.}
    \label{fig:overview:architecture}
    \vspace{-0.2in}
\end{figure}

To address these challenges, we design and build a distributed LLM serving system, \textit{\sysname}, to fully unleash the potential of ESP. Figure~\ref{fig:overview:architecture} shows the architecture of \sysname. \sysname consists of a set of elastic instances and a global manager. These elastic instances can dynamically organize themselves into a set of disjoint ESP groups to process batches of requests in parallel with different configurable degrees of parallelism. They can also support efficient elastic scaling up and down without additional overhead to adapt to the varying resource demands of requests (\S\ref{sec:design:mechanism}). The GPU memory of \sysname elastic instances also forms a unified distributed key-value cache pool, which can be flexibly used to store key-value tensors of requests at the granularity of a token across elastic instances to reduce GPU memory fragmentation (\S\ref{sec:design:mechanism}). The \sysname global manager is responsible for managing requests, elastic instances, and the unified distributed key-value cache pool with the global view.
 
In each iteration, based on profiling results from the scaling information base (SIB), the global manager dynamically adjusts the DoP of existing batches, grouping strategy of elastic instances, batching and dispatching strategy of newly arrived requests, and placement strategy of key-value caches to improve the throughput and reduce the latency of requests in real-time (\S\ref{sec:design:algorithm}). In this process, the \sysname dispatcher accompanied by the global manager dispatches newly arrived requests to a set of specific elastic instances as the global manager requires. At the same time, given the DoP and scaling plan generated by the global manager, the elasticity controller orders elastic instances to update their configuration to form the corresponding ESP groups and process requests in parallel. The key-value caches generated from LLM inference are placed at the position as specified in the scaling plan. The global manager monitors the progress of requests, the resource usage of elastic instances, and the key-value cache pool to update its decision.
\section{\sysname Elastic Instances}
\label{sec:design:mechanism}

Elastic instances are the minimum independent execution units in \sysname. Each maintains a replica of the model weights and employs a unified model parallelisms on the equivalent number of GPUs (\S\ref{sec:design:algorithm:scaling_plan}). Before each iteration, the global manager dynamically assigns elastic instances into multiple parallel groups, where each group handles the computation for a specific batch with different degrees of parallelism (DoP). The number of instances in a parallel group affects its corresponding DoP. Elastic instances may also receive potential elastic scaling plans from the global manager, which entails the reassignment of instances' parallel groups. If so, upon completion of the iteration's computation, instances are required to realize elastic scaling to ensure that subsequent computations can proceed according to the newly defined DoP, meaning that the necessary key-value tensors are already in each instances's KV Cache pool.

\begin{figure}[t]
    \includegraphics[width=\linewidth]{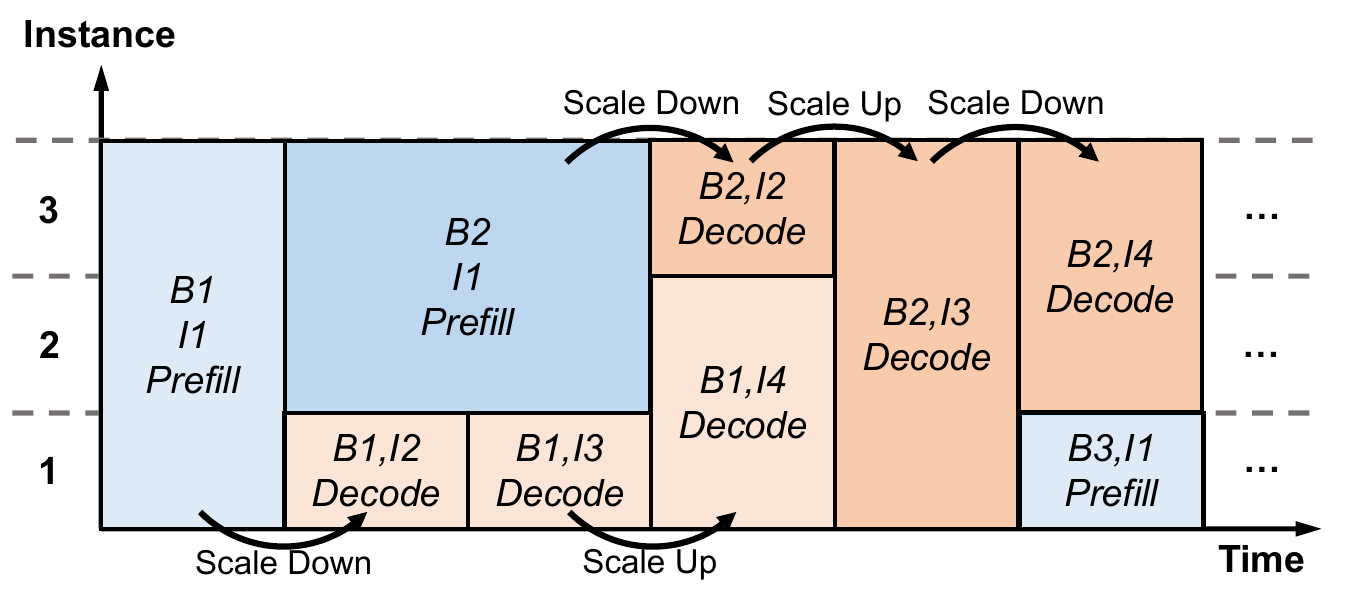}
    \vspace{-0.3in}
    \caption{Lifecycle of requests.}
    \label{fig:design:lifecycle}
    \vspace{-0.27in}
\end{figure}

Figure~\ref{fig:design:lifecycle} shows the lifecycle of requests, where $(B_i, I_j)$ indicates the $j$-th iteration of batch $i$. Because the computational complexity of the prefill phase is much higher than that of the decoding phase, it is always necessary to scale down the batch after the prefill phase ($(B_1, I_1)$ and $(B_2, I_1)$). As the generated tokens become more and more, the computational complexity of the decoding phase increases and the GPU memory in a parallel group may be filled up with the key-value cache, leading to the necessity of scaling up the parallel group ($(B_1, I_3)$ and $(B_2, I_2)$). It can also optionally scale down the decoding batch to leave more resources for the prefill batch ($(B_2, I_3)$).

To mitigate the overhead of scaling mechanisms, we have design a set of zero-overhead ESP mechanisms, which allow for elastic scaling down in the prefill phase and scaling up in the decoding phase without any additional key-value tensor migrations. In this section, we elaborate on the design of these mechanisms. As for the optional scaling down in the decoding phase, the global manager only uses it when its benefits outweigh the overhead of scaling down (\S\ref{sec:design:algorithm}).

\subsection{Elastic Scale-down}

After a batch completes the prefill phase, its computational demand for its decoding phase decreases significantly. In this case, it is usually beneficial to scale down the size of its parallel group to release resources for other batches. Specifically, for a parallel group $R$ with DoP $d$, the elastic scale-down mechanism needs to scale down the group to a new parallel group $R'$ with DoP $d'$, where $d' < d$. The primary challenge is to ensure that the entire key-value tensors of requests in the parallel group $R$ are efficiently transferred to the new parallel group $R'$.

\parabf{Existing solution: reactive migration.}
Existing practices~\cite{patel2023splitwise, zhong2024distserve, hu2024inference} use reactive migration that migrates the key-value tensors from the parallel group $R$ to the new parallel group $R'$ after the prefill phase. However, the migration overhead is non-negligible and escalates linearly with the sequence length of requests. Specifically, for a request with an input length of 1M, the GPU consumption of the key-value tensors exceeds 488GB when serving a 7B LLM. Even if GPUs are connected by high-bandwidth interconnects, such as NVLINK and Infiniband, migrating a single request still spans several seconds, significantly longer than a decoding step.

Additionally, reactive migration suffers from the GPU fragmentation problem. It requires at least $O(blsh/d)$ unused GPU memory for key-value tensors in \textit{each} instance of $R$ to accommodate the newly generated key-value tensors before migration can proceed, where $b$, $l$, $s$, $h$ are batch size, layer numbers, sequence length, hidden dimension, respectively. For instance, consider a request with an input length of 600K, and the DoP $d$ is set to 3. When the unused slots in three instances are 100k, 200k, and 400k, respectively, the parallel group cannot handle the request, even if the total number of unused key-value slots suffice. The reason is that reactive migration first generates 200K key-value tensors on each instance, leading to an Out of Memory (OOM) error on the first instance.

A potential mitigation method is to distribute request tokens unevenly across instances based on their available unused key-value cache slots, rather than equally. However, this uneven distribution may lead to severe computational imbalances across instances, significantly delaying the entire batch's processing time, because the prefill phase is bottlecked by the slowest instance.

\parabf{Our solution: proactive migration.}
To eliminate the migration overhead, we propose a new scale-down mechanism without additional communication overhead, called proactive migration. Our key observation is that during the prefill phase, sequence parallelism inherently circulates the key-value tensors among the parallel group. Instead of reactive migration after the prefill phase, we selectively retain key-value tensors in the KV cache pool of instances in the new parallel group $R'$ during the prefill phase, thus achieving zero-overhead elastic scaling down.

\begin{figure}[t]
    \includegraphics[width=\linewidth]{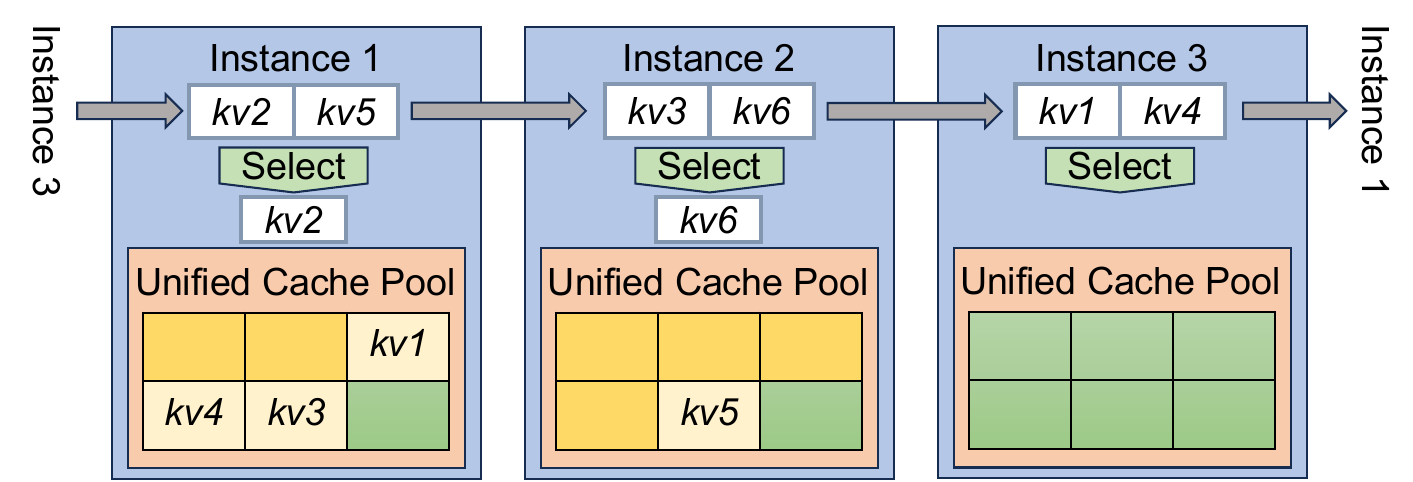}
    \vspace{-0.2in}
    \caption{Elastic scale down in the prefill phase.}
    \label{fig:design:elastic_context_phase}
    \vspace{-0.2in}
\end{figure}

Figure~\ref{fig:design:elastic_context_phase} shows an example. In this example, three instances form a parallel group $R$ with DoP=3, and they are instructed to scale down to a new parallel group $R'$ consisting of the first two instances, where the first four tokens stored in instance 1, and the rest tokens stored in instance 2. In this case, when receiving key-value tensors from the neighboring instance, besides computing the attention and sending them to the next instance, the first two instances also selectively save key-value tensors into their key-value cache pool as the instruction requires. After the computation of the prefill phase, the key-value tensors of requests are already in the KV cache pool of $R'$.

Compared to reactive migration, proactive migration incurs no additional migration overhead, because it reuses existing communication results. Furthermore, proactive migration eliminates memory constraints imposed by reactive migration and allows for any token-level KV Cache allocation plan according to the memory availability of each instance without computational load imbalance. Besides, it also reuses existing buffer space in the sequence parallelism, avoiding additional GPU memory allocation. Because the computation is conducted layer-by-layer, this buffer only needs to temporarily stores a single layer's key-value tensors. The buffer size, i.e., $O(bsh/d)$, may be even smaller than that of the pure model parallelism, which requires $O(bsh)$.

\subsection{Elastic Scale-up}
Due to the decoding phase's lightweight computation pattern, the decoding batch often executes with a small DoP for overall efficiency. However, as the decoding progresses, the size of the generated key-value tensors may exceed the capacity of the KV cache pool in the parallel group, necessitating adding new instances to expand the capacity. Moreover, if the batch size is sufficiently large, the computation in the decoding phase may become compute-bound. In such cases, employing more GPUs can reduce latency, but also require an efficient scale-up operation.
When scaling up the parallel group, the primary challenge is to ensure that the newly added instances can efficiently participate in the ongoing computation without incurring additional overhead.

\parabf{Existing solution: entire requests migration.} Previous works such as tensor-parallelism~\cite{shoeybi2020megatronlm} (TP) and FlashDecoding~\cite{dao2023flashattention2} only support distributed decoding computations across multiple GPUs within a \textit{single} instance. When resources of an instance, such as GPU memory, are insufficient, they have to migrate some requests in the batch to another instance, and then parallelly process them in different instances. However, migrating entire key-value caches of a request to another instance incurs significant migration overhead, even higher than a decoding step itself. Additionally, it requires the entire or most of the key-value tensors of a request must be stored in a single instance, leading to memory fragmentation issues.

\parabf{Our solution: multi-master distributed decoding.} We first extend sequence parallelism to the decoding phase, called single-master distributed decoding, allowing multiple instances to participate in the decoding computation of a single batch. In a parallel group, an instance is designated as the master instance, responsible for driving the computation process. At each attention layer, it first computes query and key-value tensors, saves key-value tensors into its local KV-Cache pool, and sends query tensors to other instances in the parallel group. All instances execute the local attention computation in parallel and send results back to the master instance. After that, the master instance proceeds to compute other local layers, such as the FFN layer, and starts to execute the next attention layer. As long as a single instance has enough memory to store newly generated key-value tensors, the entire batch can be processed across multiple instances by designating it as the master instance. When scaling up the parallel group, the global manager only needs to add new instances to the parallel group and instruct them to execute without migrating existing key-value tensors.

\begin{figure}[t]
    \includegraphics[width=0.6\linewidth]{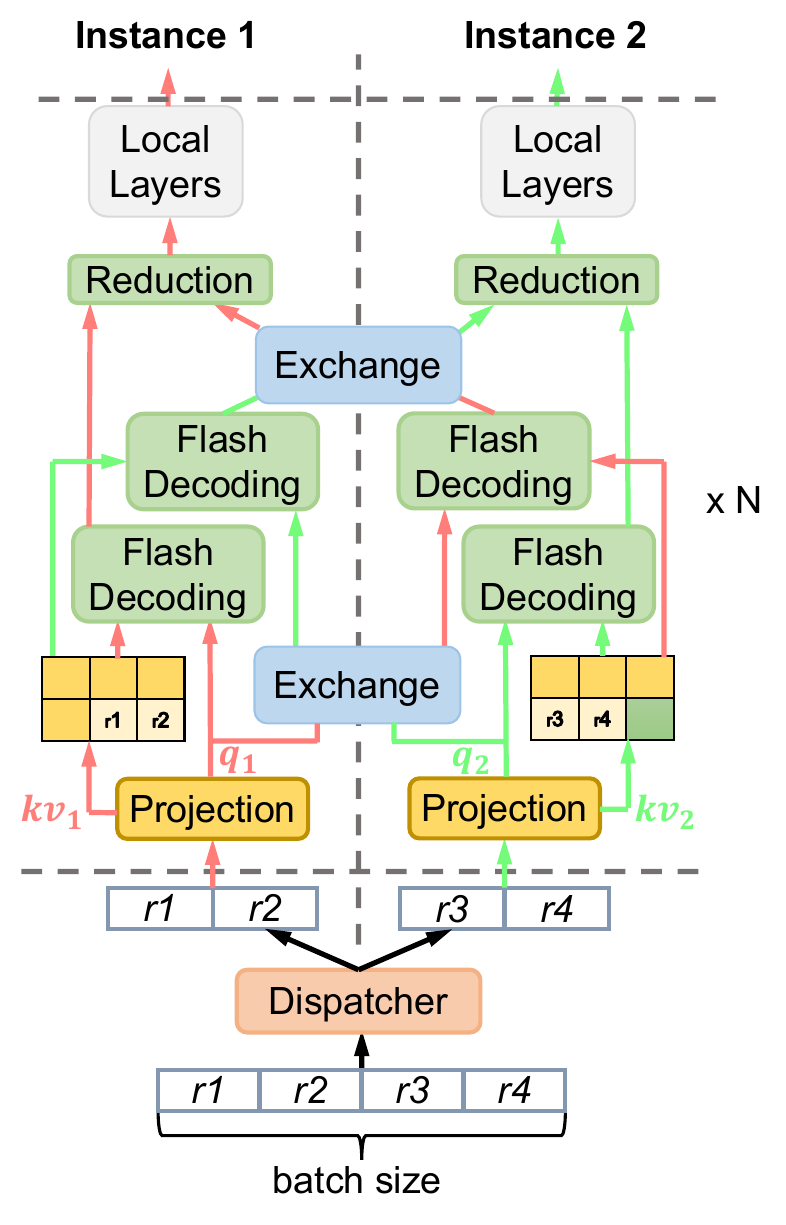}
    \vspace{-0.2in}
    \caption{Elastic scale up in the decoding phase.}
    \label{fig:design:elastic_decoding_phase}
    \vspace{-0.3in}
\end{figure}

However, this approach has its limitations when the batch size is large. In terms of memory management, it requires that the unused slots in the master instance are sufficient to store key-value tensors generated in the next iteration, leading to memory fragmentation issues. In terms of computation, since the local layers like FFN are all performed on the master instance, when the decoding phase becomes compute-bound, its performance is restricted by the computation resources in the master instance.

To address these issues, we further extend it to multi-master distributed decoding. As shown in Figure~\ref{fig:design:elastic_decoding_phase}, a parallel group contains multiple master instances. Different master instances are responsible for different requests in a batch. They save corresponding key-value tensors into their local KV cache pools to break the memory fragmentation issue, and parallelize local layer computations across multiple master instances to improve computational efficiency. When master instances exchange query tensors, the communication between them can further overlap with the local attention computation of its mastered requests.

\section{\sysname Global Manager}
\label{sec:design:algorithm}

\begin{figure}[t]
    \includegraphics[width=\linewidth]{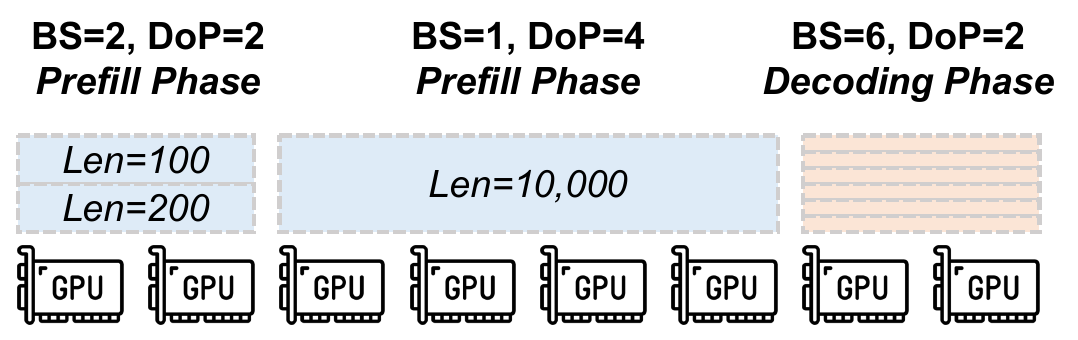}
    \vspace{-0.2in}
    \caption{Illustration of the flexible scheduling space.}
    \label{fig:design:scheduling_space}
    \vspace{-0.3in}
\end{figure}

With efficient elastic scaling mechanisms to change the DoP, the global manager is responsible for using them to schedule requests and key-value tensors across elastic instances efficiently. There may be newly arrived requests in the pending queue $P = \{r_1, r_2, ..., r_{n_P}\}$, a set of batches of requests in the decoding phase $B = \{B_1, B_2, ..., B_{n_B}\}$, where each decoding batch $B_i$ is associated with an existing parallel group $G_i$. Elastic instances are either idle due to scale-down operation in the last iteration or executing a decoding batch. In each iteration, the global manager needs to dispatch some requests from $P$ to execute in the prefill phase, allocate elastic instances to them, decide the batching strategy, and change the states of existing decoding batches. As shown in Figure~\ref{fig:design:scheduling_space}, there are lots of requests with different sequence lengths in the different phases. All these aspects, including batching strategy, the DoP of each batch, and placement of key-value tensors, are configurable and may be different across iterations. It forms a scheduling space exponential to the number of requests and elastic instances.

The primary challenge for the global manager is to generate an efficient scheduling plan from this complex scheduling space in real time. The real-time requirement is due to the dynamic nature of the LLM inference workload. The efficiency of a scheduling plan is highly dependent on the current workload. Different workload requires different scheduling plans. For example, when serving a request with sub-linear scalability under the light load, as long as scaling-up is beneficial, it is better to serve this request with a high DoP to use more idle GPUs for GPU utilization improvement. Conversely, if under heavy load, due to the request's poor scalability, it is better to serve this request with a low DoP to leave more resources for other requests for GPU utilization improvement. It indicates that the global manager needs to make decisions in real-time based on the real-time state. However, the LLM inference is super fast. The duration of an iteration can be as low as tens of milliseconds. The scheduling time is restricted by a limited time budget.

To address these issues, we propose a \textit{scalable four-step scheduling algorithm}. The key insight is that we decouple this scheduling problem into four sub-problems: dispatching, elastic instance allocation, batching, and elastic scaling plan generation. In each step, the \sysname global manager generates an efficient plan for a respective aspect in polynomial time. Then, the \sysname global manager combines these plans to form the final scheduling plan.

\subsection{Dispatching}
\label{sec:design:algorithm:dispatching}

The dispatching step is to choose a subset of requests $R_p$ from the pending queue $P$ to execute the prefill phase in the current iteration. In this step, the global manager considers resource availability in both GPU computing and GPU memory. As in prior works~\cite{yu2022orca, kwon2023efficient}, the global manager scans through $P$ in the first-come-first-serve (FCFS) order.

\parabf{GPU memory constraints.}
As for GPU memory, GPU memory consumption of key-value tensors is the primary concern. The global manager does not add a request into $R_p$ unless there are sufficient unused slots for this request. Moreover, because evicting a long sequence request incurs significant recomputation overhead, the global manager also tries to avoid future eviction and recomputation due to insufficient GPU memory. As in prior works~\cite{wu2023fast}, the global manager estimates it by considering the maximum future key-value cache consumption based on the current state and the maximum sequence length of the request given by users. The global manager does not add it into $R_p$ if it may trigger eviction.

\parabf{GPU computing constraints.}
As for GPU computing, the primary concern is the efficiency of dispatching $R_p$ and their impact on existing requests in the decoding phase $B$. The global manager uses an analytical model and profiling results from SIB (\S\ref{sec:design:algorithm:optimization:analytical_model}) to measure them in terms of iteration time $T(R_p, E_p)$ on the occupied instances $E_p$. $E_p$ is initialized as the subset of all idle instances.

First, there is a tipping point that $R_p$ transitions from memory bound to compute bound. Before that point, adding more requests into $R_p$ improves the efficiency of GPU computing. After that point, it only extends the execution time with negligible efficiency improvement. We estimate this point by profiling the upper bound of the iteration time that a prefill batch is memory bound. The global manager stops adding more requests into $R_p$ when the iteration time of $R_p$ exceeds this tipping point.

Second, adding more requests into $R_p$ might interfere with some existing batches $B_p = \{B_{p, 1}, B_{p, 2}, ...\} \subseteq B$ if new requests utilize some unused key-value cache slots in instances occupied by $B$. To simplify this problem, we conservatively consider it in the worst case that $R_p$ may preempt $B_p$. In this case, for each $B_{p, i}$, unused key-value slots of instances in its parallel group $G_{p, i}$ can be used to add an additional subset of new requests $R_{p, i}'$ into $R_p$. The global manager analyzes the performance gain of executing $R_{p, i}'$ and the cost of preempting $B_{p, i}$ to decide whether to add $R_{p, i}'$ into $R_p$ and expand $E_p$. The performance cost, i.e., the impact of preemption on the output token latency, is formulated as follows:
\begin{equation}
    \text{Cost} = \sum_{r \in B_{p, i}} \frac{T(R_p \cup R_{p, i}', E_p \cup G_{p, i})}{r.\text{output\_len}}
\end{equation}
In this equation, the output length of requests is the number of existing output tokens. This estimation is similar to the output length-based waiting time estimation in previous LLM serving works~\cite{wu2024dlora, smartspec, patke2024one}.
As for the performance gain, it is formulated as the impact on the input token latency of $R_{p, i}'$:
\begin{equation}
    \text{Gain} = \sum_{r \in R_{p, i}'} \frac{(\text{AvgLat}_{\text{d}} - \min(B_{p, i}.\text{exec\_time}))^+}{r.\text{input\_len}}
\end{equation}
In this equation, $\text{AvgLat}_{\text{d}}$ is the average execution time of finished requests in the decoding phase, and $\min(B_{p, i}.\text{exec\_time})$ is the executed time of requests in the decoding phase. The subtraction between them estimates how long $R_{p, i}'$ have to wait for the decoding batch in the worst case. The global manager adds $R_{p, i}'$ into $R_p$ and adds $G_{p, i}$ into $E_p$ if the gain is greater than the cost. The complexity of this step is $O(n_B)$.

\subsection{Elastic Instance Allocation}
After generating exact $R_p$ to execute, the global manager needs to decide actual elastic instance allocation for them in this step. In this step, the primary concern is to mitigate the interference between the prefill phase, i.e., $R_p$, and the decoding phase, i.e., $B$, while maximizing the efficiency of GPU computing.

The global manager first allocates idle instances to $R_p$. If the unused key-value cache slots of idle instances are not enough, $R_p$ can preempt a few instances with the most unused key-value cache slots to obtain sufficient key-value cache slots. To avoid preemption, the global manager tries to migrate existing key-value tensors in preempted instances to other active instances if possible. As a result, $R_p$ can obtain elastic instances $E_p$ with sufficient key-value cache slots.

However, it is still possible to achieve better performance by allocating more elastic instances to the compute-intensive prefill phase. To this end, the global manager repeatedly considers whether to allocate elastic instances with the fewest used key-value cache slots, $e_{min}$, to $R_p$. The constraint is that the batch using $e_{min}$ can migrate its key-value tensors to other instances in the decoding phase. In this case, the input token latency is reduced as follows:
\begin{equation}
    \text{Gain} = \sum_{r \in R_p} \frac{T(R_p, E_p) - T(R_p, E_p \cup {e_{min}})}{r.\text{input\_len}}
\end{equation}
But it also incurs the migration overhead as follows:
\begin{equation}
    \text{Cost} =  \sum_{r \in R_p} \frac{V(e_{min})}{\text{avg\_bandwidth} \cdot r.\text{input\_len}}
\end{equation}
In this equation, $V(e_{min})$ is the volume of existing key-value tensors in $e_{min}$, and the avg\_bandwidth is the average bandwidth between $e_{min}$ and target instances. Target instances are always instances with the most unused key-value cache slots. The global manager repeatedly allocates $e_{min}$ to $R_p$ until the gain is less than the cost. The complexity of this step is $O(m)$, where $m$ is the number of elastic instances.

\subsection{Batching}
After deciding $R_p$ and $E_p$, this step optimizes the batching strategy of $R_p$ on $E_p$ to further minimize the latency of the prefill phase.  In this step, the primary concern is to assign different DoPs to requests with different sequence lengths.

To address this issue, we formulate this batching problem as a dynamic programming (DP) problem. The optimization goal is to minimize the input latency of $R_p$ on $E_p$. Our key insight is that requests with similar sequence lengths have similar characteristics and should be batched together. Therefore, the global manager first sorts requests based on their sequence lengths in descending order. The allocated elastic instances are also sorted based on their locations and the number of unused key-value cache slots in ascending order. Let $f[i][k]$ be the minimum input latency of the first $i$ requests when using the first $k$ elastic instances. The DP equation can be formulated as follows:

\begin{equation}
    f[i][k] = \min_{0 < j \leq i, 0 < l \leq k, \atop D[j, i] \leq V[l, k]} \left(f[j][l] + T(R[j, i], E[l, k])\right)
\end{equation}

The $D[j, i]$ in the equation is the number of tokens of requests from $j$ to $i$, and the $V[l, k]$ is the number of unused key-value cache slots of elastic instances from $l$ to $k$. These sums of values in an interval can be calculated by maintaining a prefix sum array in advance. The $T(R[j, i], E[l, k])$ is the sum of input latency of requests from $j$ to $i$ when using elastic instances from $l$ to $k$. The minimum input latency sum of all requests, i.e., $\min_{0 < j \leq m} f[n][j]$, can be found in polynomial time. When updating $f[i][k]$, the global manager records the last split point of requests $split_{req}[i][k]$, i.e., $j$, and the last split point of elastic instances $split_{ins}[i][k]$, i.e., $l$. The global manager uses them to backtrack to generate the batching plan and the corresponding DoP for each batch.

If naively update $f[i][k]$ for all $0 < i \leq n$ and $0 < j \leq m$, the time complexity of this DP algorithm is $O(|R_p|^2 \cdot |E_p|^2)$. However, we notice that $split_{req}[i][k]$ and $split_{ins}[i][k]$ have following properties:

\begin{equation}
\begin{aligned}
    split_{req}[i][k-1] \leq split_{req}[i][k] \leq split_{req}[i][k+1],
    \\
    split_{ins}[i-1][k] \leq split_{ins}[i][k] \leq split_{ins}[i+1][k].
\end{aligned}
\end{equation}

Therefore, this problem can be optimized to $O((|R_p|+|E_p|)^2)$ by using \textit{Quadrangle Inequality Properties}~\cite{quadrangle}. Although it can further optimize the time complexity, it is efficient enough in practice.

\subsection{Elastic Scaling Plan Generation}
\label{sec:design:algorithm:scaling_plan}
As described in \S\ref{sec:design:mechanism}, the global manager also needs to generate elastic scaling plans for proactive scaling down and up. 

For proactive scaling down, the key insight is that the decoding phase scales poorly. As shown in Figure~\ref{fig:motivation:scaling_curves}, in most cases, the minimum best DoP of requests in the decoding phase are similar and are smaller than that in other cases. Therefore, we set the degree of model parallelism as the minimum best DoP at launch time. At run time, the global manager only needs to scale down the DoP to the minimum DoP that the key-value tensors of requests can fit in the corresponding elastic instances. It is optimal for most requests in the decoding phase. Even for requests in the decoding phase with larger best DoPs, it is still near-optimal, because leaving more elastic instances to the compute-intensive prefill phase with longer duration is more beneficial.

For scaling up, the global manager scales up when GPU computing or GPU memory is insufficient. Insufficient GPU computing refers to the decoding phase becoming compute-bound. Because FFN layers first become the computation bottleneck and their complexity is related to the batch size, the global manager uses a batch size threshold to detect it. We profile this threshold in advance. The multi-master decoding is used as long as it can reduce memory fragmentation or execution time. The number of newly key-value tensors generated by each master is set to as uniform as possible.

\subsection{Optimizations}

\parabf{Analytical model based on SIB.}
\label{sec:design:algorithm:optimization:analytical_model}
The iteration time of the prefill phase under different scenarios guides the decision-making. Ideally, we can record them into SIB in advance and retrieve them at run time. However, there are massive combinations of requests with different input lengths executing at different DoPs. It is impossible to cover all the cases only by profiling. Therefore, we propose an analytical model to estimate them, which is formulated as follows:
\begin{equation}
    \text{T}_{p}(R) = \alpha_p + \beta_p \cdot \sum_{r \in R} r.\text{input\_len} + \gamma_p \cdot \sum_{r \in R} r.\text{input\_len}^2
\end{equation}
In this equation, $\alpha_p$, $\beta_p$, and $\gamma_p$ are the coefficients capturing the constant overhead, linear computation (e.g., FFN layers), and quadratic computation (e.g. attention layers), respectively. They are trained by the least square method based on a few profiling results. For different parallelism strategies, we train different coefficients.

\section{Implementation}
\label{sec:implementation}

\sysname is implemented in approximately 15K lines of code based on C++, CUDA, Python, and Triton~\cite{triton}, and reuses some components from vLLM~\cite{kwon2023efficient} and LightLLM~\cite{lightllm}. The front end of \sysname is similar to OpenAI API~\cite{openai2023gpt4}. Users send requests to \sysname based on the front-end API. The code of \sysname is open-source and is publicly available at
\url{https://github.com/LoongServe/LoongServe}.

The \sysname global manager is mainly implemented in Python, but some core logic, such as the batching algorithm, is implemented in C++ to accelerate looped functions. To manage multiple batches at the same time, the global manager assigns each batch to a Python coroutine.

The global manager uses Ray~\cite{moritz2018ray} to communicate with elastic instances. Because ESP introduces extra RPC parameters, RPC parameters is carefully designed to reduce extra serialization overhead. Elastic instances also cache active ESP metadata. Similar to vLLM~\cite{kwon2023efficient}, when tensor parallelism is enabled, the global manager mainly sends information to a single rank in an elastic instance, and this rank uses NCCL~\cite{nccl} to broadcast the information to other ranks to further reduce serialization overhead.

\begin{figure*}[t]
    \includegraphics[width=\linewidth]{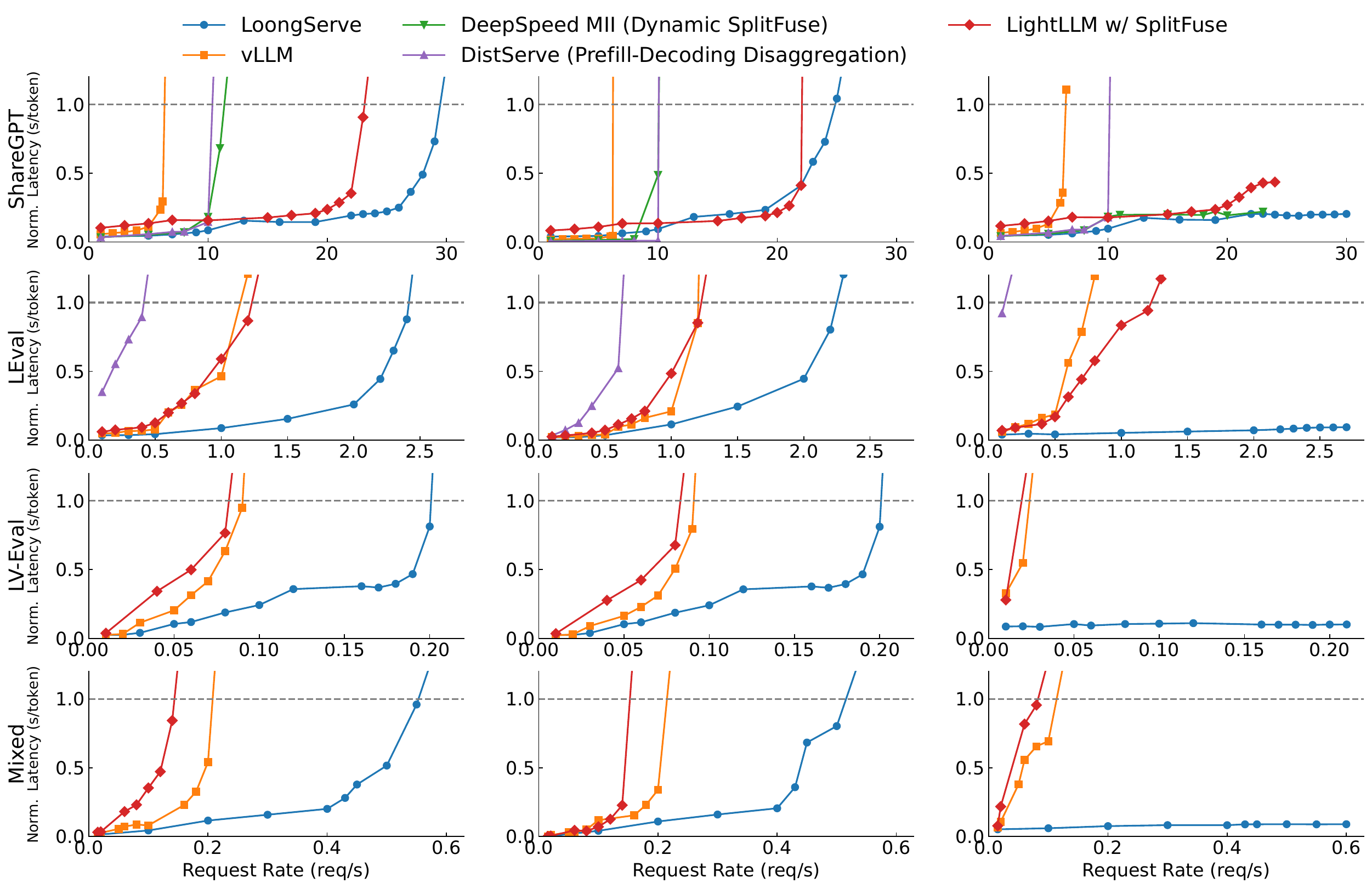}
    \hspace*{0.3in}{Avg. per-token latency.}\hspace*{\dimexpr\linewidth/8\relax}{Avg. input token latency.}\hspace*{\dimexpr\linewidth/8\relax}{Avg. output token latency.}
    \caption{Average latency of different LLM serving systems with the LWM-1M-Text (Llama-2-7B) on real-world workloads.}
    \label{fig:evaluation:e2e}
\end{figure*}

For each elastic instance, it manages its corresponding key-value cache pool by using PagedAttention~\cite{kwon2023efficient} at the granularity of a single token. Although StripedAttention performs pretty well on long sequences, it causes redundant computation in short sequences due to its special causal attention mask. We tune the tile size to skip most redundant computations in short sequences. We also reduce the lifecycle of extra parameters introduced by StripedAttention in shared memory to improve the occupancy of streaming multiprocessors (SMs). For the decoding phase, we implement a custom version of Flash-Decoding~\cite{dao2023flashattention2} with extra parameters to support ESP. All the above optimizations are compatible with MHA, MQA, and GQA, and have the same accuracy as the original implementations.

The communication between elastic instances is based on NCCL with dedicated CUDA streams. Tensor parallelism and sequence parallelism use different NCCL communicators. To support multiple dynamic parallel groups at the iteration level, we use NCCL group functions to merge multiple point-to-point operations to form collective operations in selected NCCL ranks.

\sysname also provides tools to generate profiling results under different scenarios. These profiling results are stored in a SQLite database. Each time training analytical models just needs to select the corresponding profiling results from the database.
\section{Evaluation}
\label{sec:evaluation}

In this section, we evaluate the performance of \sysname with state-of-the-art solutions on different real-world workloads and show the effectiveness of its components.

\subsection{Methodology}
\label{sec:evaluation:setup}

\parabf{Model.}
We use the LWM-1M-Text model~\cite{liu2023world} as the long-context LLM model in our evaluation. It is an open-source pre-trained LLM with the largest context window size (1 million tokens) when we started our evaluation. Besides, it uses the same model architecture as Llama-2-7B~\cite{touvron2023llama2}, which is widely used in practice.

\parabf{Testbed.}
We evaluate \sysname on servers each with eight NVIDIA A800 80GB GPUs, 128 CPUs, 2048 GB of host memory, and four 200 Gbps InfiniBand NICs. The NVLink bandwidth between two GPUs is 400 GB/s. We use PyTorch 2.0.0, CUDA 12.2, OpenAI Triton 2.1.0, and HuggingFace tokenizers 0.15.2 for our evaluation. Most experiments are conducted on a single server, and we also evaluate the multi-node performance of \sysname on two servers.

\parabf{Workloads.} Similar to prior work~\cite{li2023alpaserve,kwon2023efficient,zhong2024distserve}, the arrival pattern of requests is generated by a Poisson process. The input lengths and output lengths of requests are sampled from the following real-world datasets.
\begin{itemize}[leftmargin=*]
    \item \textbf{ShareGPT}~\cite{sharegpt}: It is collected from real-world conversations with ChatGPT and is widely used in prior work~\cite{kwon2023efficient,wu2023fast,zhong2024distserve}. Due to the limited context window of ChatGPT-3.5, the range of sequence length in this dataset is 4 - 2.3K tokens.
    \item \textbf{L-Eval}~\cite{an2023leval}: It contains human-labeled query-response pairs on diverse tasks, such as summarization and question answering. It is used to evaluate the long-context capability of Qwen 1.5~\cite{qwen}. The range of sequence length in this dataset is 2.7K - 210.5K tokens.
    \item \textbf{LV-Eval}~\cite{yuan2024lveval}: It is the dataset containing the longest requests when we started our evaluation. It contains long-context question-answer tasks. When converting words into tokens, the range of sequence length in this dataset is 15.1K - 497.3K tokens.
    \item \textbf{Mixed}: Because the above datasets only cover a small range of sequence lengths, we also mix them to evaluate the performance of \sysname on a more diverse workload. The sampling probability of each dataset is the same.
\end{itemize}

\begin{figure}[t]
    \includegraphics[width=\linewidth]{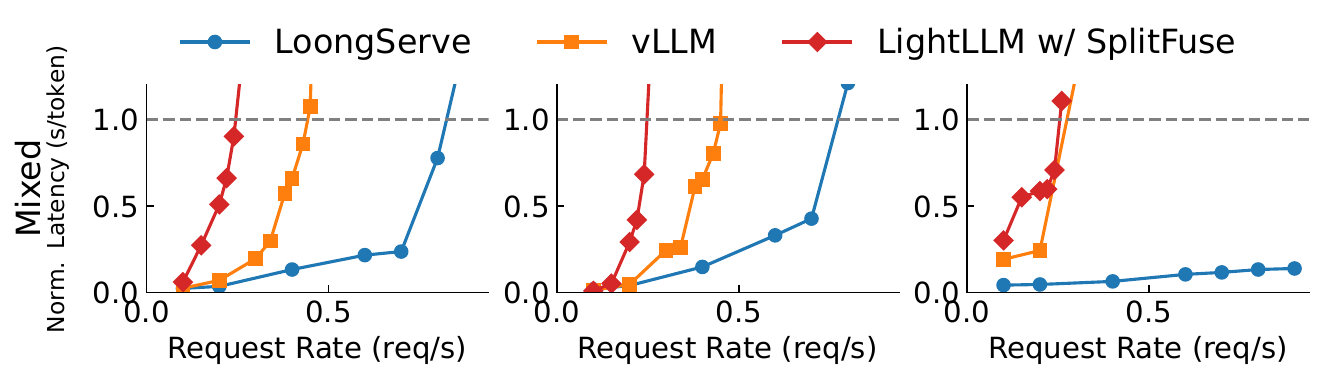}
    \vspace{-0.05in}
    \hspace*{0.1in}{Per-token latency.}\hspace*{\dimexpr\linewidth/16\relax}{Input latency.}\hspace*{\dimexpr\linewidth/16\relax}{Output latency.}
    \vspace{-0.1in}
    \caption{Multi-node performance.}
    \label{fig:evaluation:e2e_multi_node}
    \vspace{-0.25in}
\end{figure}

\parabf{Baselines.}
We compare \sysname with the following state-of-the-art LLM serving systems:
\begin{itemize}[leftmargin=*]
    \item \textbf{vLLM}~\cite{kwon2023efficient}\footnote{vLLM 0.3.0, commit hash: 1af090b57d0e23d268e79941f8084bf0a8ad8621}: It is one of the most popular LLM serving systems. To fully leverage all GPUs and serve requests with long context, we set the tensor parallelism to 8.
    \item \textbf{DeepSpeed-MII}~\cite{deepspeedmii}\footnote{DeepSpeed MII, https://github.com/microsoft/DeepSpeed-MII, commit hash: 773b735d6294a98dd842d82ef024d0d9b050f66aa
    }: It proposes Dynamic SplitFuse~\cite{holmes2024deepspeedfastgen} to decompose long input sequences into small chunks to prevent performance degradation of the decoding phase when serving both phases. However, when serving requests with long context larger than 32K tokens, we encounter the "illegal memory access" error, so we only evaluate it on ShareGPT. The tensor parallelism is set to 8.
    \item \textbf{LightLLM w/ SplitFuse}~\cite{lightllm}\footnote{DistServe, https://github.com/LLMServe/DistServe, commit hash: e2b5168a50f24d960ead314b0649428e35381f80
    }: To evaluate the performance of SplitFuse for long sequences, we also evaluate SplitFuse provided by LightLLM. Similar to SARATHI~\cite{agrawal2023sarathi}, it needs to set the chunk size. We set the chunk size as the ideal "P:D ratio" described in SARATHI~\cite{agrawal2023sarathi} by calculating it for each dataset before experiments, although it is unknown in practice. The tensor parallelism is set to 8.
    \item \textbf{DistServe}~\cite{zhong2024distserve}: It proposes prefill-decode disaggregation to reduce the impact of the prefill phase on the decoding phase. We contact the authors to get the source code and set parallelism strategies for it. Because its parallelism strategies are restricted by the head of the model and requests with long context need large GPU memory, we use four GPUs for the prefill phase and four GPUs for the decoding phase. The DoP for each phase is set to 4, which is the best parallelism strategy searched by DistServe's simulator and is validated on the testbed.
\end{itemize}

For \sysname, we set tensor parallelism to 2 and ESP to 4. For all the systems, the number of key-value cache slots is set as much as possible to improve the throughput.

\parabf{Metrics.}
We focus on the serving throughput. For each request rate, we measure the \textit{normalized per-token latency}, i.e., the mean of requests' end-to-end latency divided by their sequence lengths, \textit{normalized input latency}, i.e., the mean of prefill phase time divided by the input lengths, and the \textit{normalized output latency}, i.e., the mean of decoding phase time divided by the output lengths, for each system. To compare different systems, we set a latency service level objective (SLO) and compare the maximum throughput under this SLO. Similar to prior work~\cite{wu2023fast, li2023alpaserve, kwon2023efficient}, we set SLO to 25$\times$ of the inference latency.

\begin{figure}[t]
    \includegraphics[width=\linewidth]{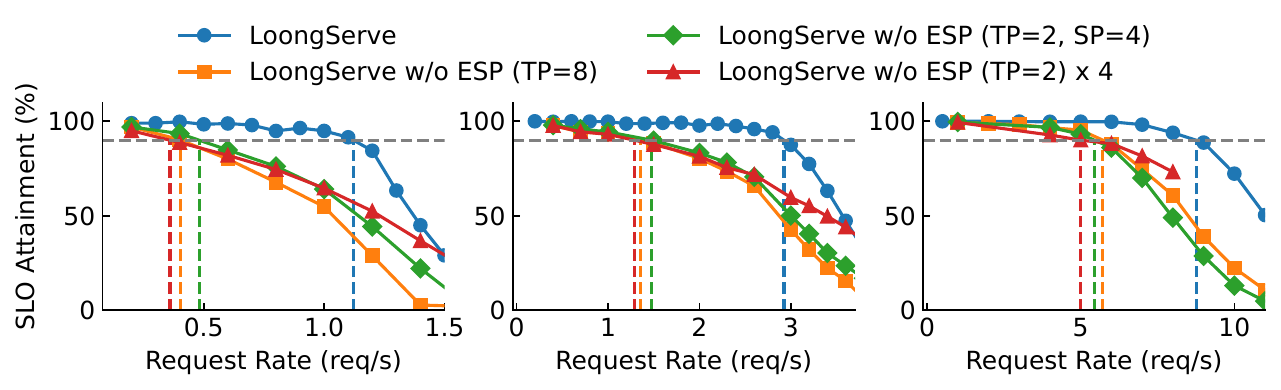}
    \vspace{-0.05in}
    \hspace*{0.2in}{(a) Zipf=1.0.}\hspace*{\dimexpr\linewidth/8\relax}{(b) Zipf=1.2.}\hspace*{\dimexpr\linewidth/8\relax}{(c) Zipf=1.4.}
    \vspace{-0.1in}
    \caption{P90 goodput under different sequence length distributions.}
    \vspace{-0.3in}
    \label{fig:evaluation:scheduling}
\end{figure}

\subsection{End-to-End Performance}

We compare \sysname with four baselines on four real-world workloads in three key metrics. Figure~\ref{fig:evaluation:e2e} shows the results. Because \sysname uses different elastic instances to execute different phases, the decoding phase is well protected from the impact of the prefill phase. As a result, the output latency of \sysname achieves low latency and is significantly better than other baselines. For the prefill phase, \sysname accelerates them by setting an appropriate DoP to avoid blocking pending prefill phases. In contrast, vLLM consistently treats different requests in the same way. On the ShareGPT, vLLM wastes resources on short requests with poor scalability. On other datasets, it causes severe interference between the decoding phase and the prefill phase. \sysname improves the total throughput and input throughput by up to 4.64$\times$ and 4.00$\times$. DeepSpeed MII and LightLLM w/ SplitFuse try to split the input sequence into chunks to protect the decoding phase. However, decomposing the input sequence makes the prefill phase inefficient. Furthermore, for long sequences in L-Eval and LV-Eval, the prefill phase still causes severe interference with the decoding phase, because the "P:D ratio" (prefill/decoding ratio)~\cite{agrawal2023sarathi} is high. Compared to them, \sysname improves the total throughput and input throughput by up to 3.85$\times$ and 3.37$\times$. DistServe disaggregates the prefill phase and the decoding phase to reduce the interference between them. However, due to the isolation between the two phases, it means that each phase can only use half of the GPUs (four GPUs) and the longest length of a request is bounded by the minimum GPU capacity of two phases, i.e., GPU memory capacity of four GPUs. On the ShareGPT, four GPUs are not enough to serve lots of requests in the decoding phase. On the L-Eval, four GPUs are not enough to serve lots of requests in the prefill phase. On the LV-Eval and Mixed, four GPUs for each phase even do not have enough memory to serve requests with long context and thus trigger the Out-of-Memory (OOM) error, so it is not shown in these two rows of figures. In contrast, with an unified distributed KV cache pool, \sysname can serve them well and achieve the best performance. Compared to DistServe, \sysname improves the total throughput and input throughput by up to 5.81$\times$ and 3.58$\times$.

\begin{figure}[t]
    \subfloat[Effectiveness of elastic scale-up.]{
        \includegraphics[width=0.48\linewidth]{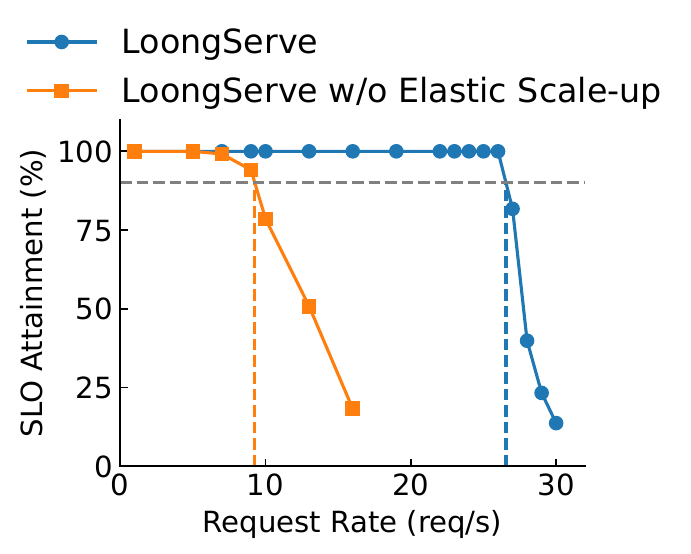}
        \label{fig:evaluation:scale-up:goodput}
    }
    \subfloat[Frequency of elastic scale-up.]{
        \includegraphics[width=0.48\linewidth]{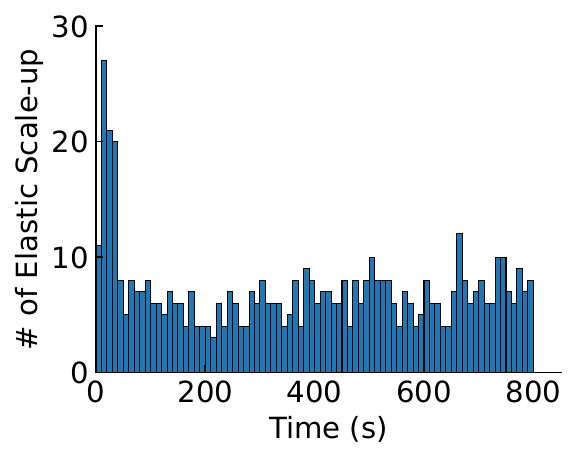}
        \label{fig:evaluation:scale-up:frequency}
    }
    \vspace{-0.1in}
    \caption{Ablation study of elastic scale-up.}
    \label{fig:evaluation:scale-up}
    \vspace{-0.2in}
\end{figure}

\subsection{Multi-Node Performance}
We also evaluate the multi-node performance of \sysname on a 16-GPU cluster. In this experiment, we deploy baselines on each server and use the same parallelism strategies as the single-node evaluation. \sysname also uses the same model parallelism strategy and extends the ESP to 8. We also use the same Mixed dataset. Figure~\ref{fig:evaluation:e2e_multi_node} shows the results. In the multi-node setting, \sysname scales well and achieves the best performance in all metrics by setting an appropriate DoP for each request. Therefore, \sysname avoids unnecessary communication overhead for short requests and enlarges the parallelism for long requests. As a result, \sysname improves the total throughput and input throughput by up to 1.86$\times$ and 1.72$\times$ compared to vLLM, 3.37$\times$ and 3.11$\times$ compared to LightLLM w/ SplitFuse, while significantly reducing the output latency under all request rates. The results show that \sysname can effectively handle the dynamic workload and achieve high performance in the multi-node setting because it can choose appropriate DoPs for diverse requests at different phases.

\subsection{Ablation Study}

\paraf{Effectiveness of elastic sequence parallelism.}
To show the effectiveness of elastic sequence parallelism, we conduct an ablation study on the P90 goodput (throughput of requests under the SLO)~\cite{zhong2024distserve} of different parallelism strategies under different sequence length distributions. Baselines include traditional tensor parallelism, i.e., \sysname w/o ESP (TP=8), static hybrid parallelism, i.e., \sysname w/o ESP (TP=2, SP=4), and parallelism with replication, i.e., \sysname w/o ESP (TP=2) $\times$ 4. To simulate different scenarios, we sample the sequence lengths from the Mixed dataset with different Zipf parameters. Because the baseline with replication is not able to serve requests with long context, we limit the maximum length of requests to 200K tokens.

Figure~\ref{fig:evaluation:scheduling} shows the results. As shown in the figure, only introducing sequence parallelism is not enough to achieve good performance. Neither static hybrid parallelism nor parallelism with replication handles the dynamic inference workload. In contrast, the scalable four-step scheduling algorithm of \sysname dynamically adjusts strategies for the dynamic workload and achieves P90 goodput improvement by 2.33$\times$, 1.98$\times$, and 1.53$\times$ under different sequence distributions. P90 goodput also indicates that ESP is beneficial for most requests.

\parabf{Elastic scale-up.}
We further evaluate the effectiveness of elastic scale-up. Similar to Figure~\ref{fig:evaluation:scheduling}, we first conduct an ablation study on the P90 goodput between \sysname with and without elastic scale-up. As shown in Figure~\ref{fig:evaluation:scale-up:goodput}, when serving the ShareGPT dataset, the P90 goodput of \sysname with elastic scale-up is 2.87$\times$ higher than \sysname without elastic scale-up. The reason is that requests from the ShareGPT dataset have a relatively short input length and long output length, which requires frequent scaling up as the output length continuously increases. As shown in Figure~\ref{fig:evaluation:scale-up}, we measure the frequency of triggering elastic scale-up when the request arrival rate is 25 requests per second on the ShareGPT dataset. Each bin in the figure represents the number of elastic scale-up operations triggered by the global manager in a 10-second interval. On average, the global manager triggers 7.12 elastic scale-up operations per 10 seconds, which indicates the necessity of elastic scale-up to handle the dynamic workload.

\begin{figure}[t]
    \includegraphics[width=\linewidth]{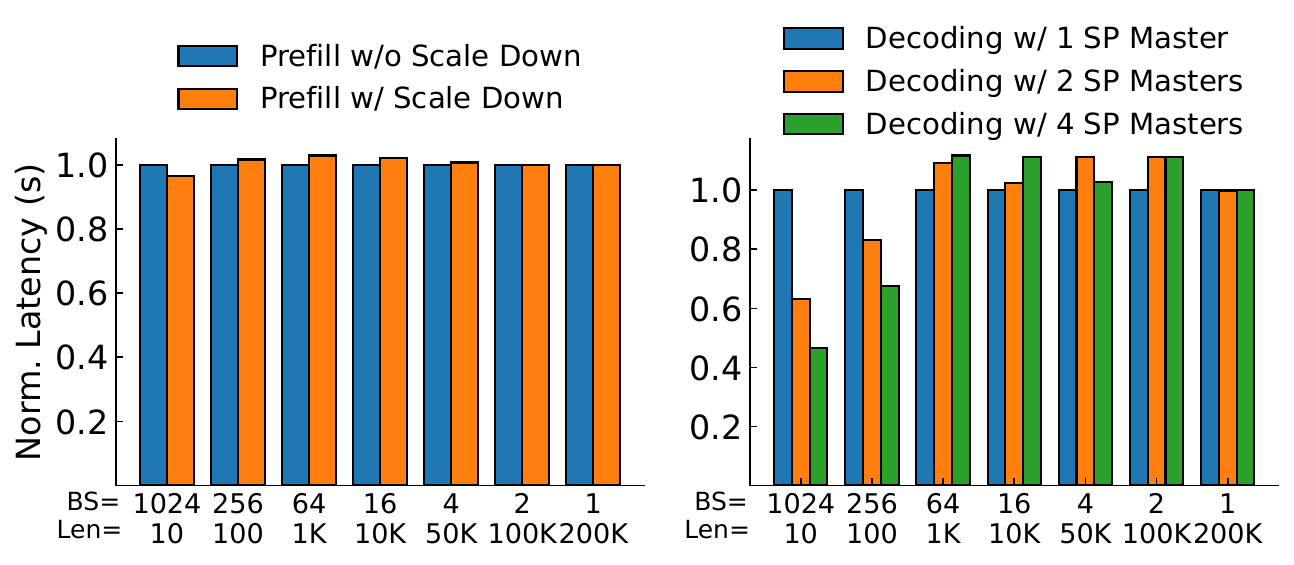}
    \vspace{-0.05in}
    \hspace*{0.2in}{(a) Scale down overhead.}\hspace*{\dimexpr\linewidth/8\relax}{(b) Scale up overhead.}
    \vspace{-0.1in}
    \caption{Overhead of elastic scaling mechanisms.
    }
    \label{fig:evaluation:ablation}
    \vspace{-0.2in}
\end{figure}

\parabf{Scaling overhead.}
We evaluate the overhead of scaling down and scaling up under different batch sizes and input lengths by timing the time to forward a batch with and without scaling, respectively. Figure~\ref{fig:evaluation:ablation} shows the results. For scaling down, we only introduce a negligible overhead (less than 2\%) for all batch sizes and prompt lengths. For scaling up, on large batch sizes, since the compute-memory ratio is high for matrix multiplication operations (including projection for query, key, and value, as well as matrix multiplication for the FFN), distributing the computation to more instances can significantly reduce the computation time, leading to a 2$\times$ improvement in per-iteration latency. On small batch sizes, the overhead of scaling up is higher because of the overhead introduced by communication and synchronization. However, the overhead is still acceptable (less than 10\%) and \sysname dynamically uses the best one.

\begin{figure}[t]
    \includegraphics[width=\linewidth]{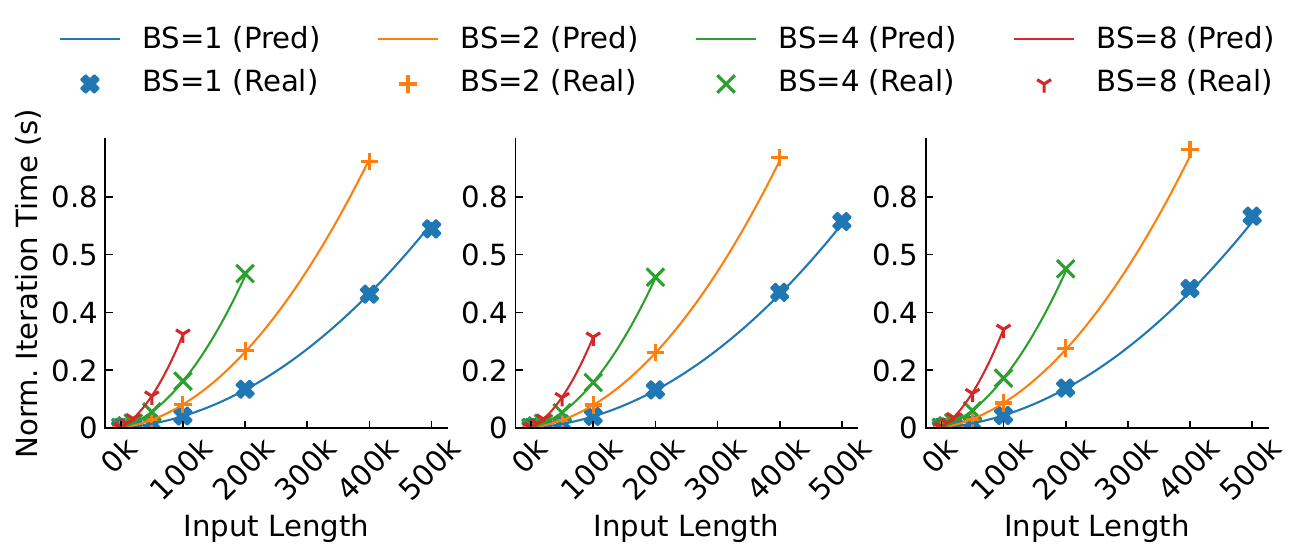}
    \hspace*{0.2in}{(a) SP2TP4.}\hspace*{\dimexpr\linewidth/8\relax}{(b) SP4TP2.}\hspace*{\dimexpr\linewidth/8\relax}{(c) SP8TP1.}
    \caption{Accuracy of \sysname analytical model.}
    \label{fig:evaluation:analytical_model_accuracy}
    \vspace{-0.15in}
\end{figure}

\parabf{Accuracy of analytical model.}
At last, we evaluate the accuracy of analytical models in different parallelism strategies. As shown in Figure~\ref{fig:evaluation:analytical_model_accuracy}, the analytical model of \sysname achieves high accuracy (less than 10\% deviation) for different batches of requests with different sequence lengths in different parallelism strategies. It is reliable to guide the \sysname global manager.
\section{Related Work}
\label{sec:related}

\parabf{LLM serving systems.}
Existing LLM serving systems adopt efficient GPU kernels for long sequence requests, such as Flash Attention~\cite{dao2022flashattention} and Flash-Decoding~\cite{dao2023flashattention2}. They are orthogonal to our work and are integrated into \sysname. To mitigate the impact of the long context, SARATHI~\cite{agrawal2023sarathi} and DeepSpeed-FastGen~\cite{holmes2024deepspeedfastgen} split the long context into chunks and process them chunks by chunks, but they still incur interference between two phases~\cite{zhong2024distserve}. SplitWise~\cite{patel2023splitwise}, DistServe~\cite{zhong2024distserve}, and TetriInfer~\cite{hu2024inference} disaggregate two phases into different groups of GPUs to avoid interference, but their static parallelism and partition strategies are not flexible to handle the dynamic workload. Infinite-LLM~\cite{lin2024infinitellm} tries to alleviate the GPU fragmentation across instances. However, it still needs periodic key-value tensor migration to maintain the locality and does not consider elastic resource demand between different requests or different phases. \sysname proposes a novel elastic scaling mechanism without additional overhead and ESP to handle dynamic workload based on requests' resource demand without locality constraints. 

\parabf{Sequence parallelism.}
Many works~\cite{liu2023ring,brandon2023striped, ulysses, li2023lightseq, korthikanti2023reducing, li2023sequence} are proposed to accelerate long-context LLM training and have demonstrated the benefit of sequence parallelism under both single-host and multi-host settings. One class of work still uses TP for the attention layers and suffers from the same problems as TP. Another class of work like Striped Attention~\cite{brandon2023striped} parallelizes the attention mechanism in the sequence dimension. All these works focus on LLM training and their DoP are fixed. Our work targets the serving scenario, supporting the decoding phase, dynamic DoP, and efficient key-value cache management.

\parabf{Long-context LLM with accuracy loss.}
Another type of Long-context LLM trades off accuracy for efficiency. Some works~\cite{child2019generating, pagliardini2023faster, sun2024linear} change the attention mechanism to reduce computation. Recent works~\cite{zhang2023h2o, xiao2024efficient, adnan2024keyformer} also try to prune the key-value cache to reduce the memory footprint.
All these works incur accuracy loss. \sysname does not affect the accuracy of the original LLM and provides a more efficient serving mechanism. Besides, \sysname is compatible with MQA~\cite{shazeer2019fast}, GQA~\cite{ainslie2023gqa}, and MoE~\cite{jiang2024mixtral} to reduce the memory footprint and computational complexity.

\parabf{Elastic training.}
Many works focus on elastic training for deep neural networks (DNNs)~\cite{li2023easyscale, elasticflow, ma2021adaptive, gu2019tiresias, pollux, afs, optimus, sia}. Compared to them, \sysname proposes a new parallelism strategy, elastic sequence parallelism, rather than using data parallelism as in the training phase. \sysname also considers more unique factors in the LLM serving, such as decoding phase, key-value tensor management and request batching, under more strict latency constraints.
\section{Conclusion}
\label{sec:conclusion}

To serve long-context LLMs under dynamic workloads, we propose ESP and build \sysname, which provides a set of elastic scaling mechanisms without additional overhead and a scalable scheduling algorithm for ESP. Evaluation across diverse real-world datasets shows that compared to existing solutions, \sysname significantly improves the performance of the prefill phase and decoding phase simultaneously.

\parabf{Acknowledgments.}
We sincerely thank our shepherd, Rong Chen, and the anonymous reviewers for their valuable feedback.
This work was supported by the National Key Research and Development Program of China under the grant number 2022YFB4500700, National Natural Science Foundation of China under the grant numbers 62172008 and 62325201.
Xin Jin is the corresponding author.
Bingyang Wu, Shengyu Liu, Yinmin Zhong, Xuanzhe Liu and Xin Jin are also with the Key Laboratory of High Confidence Software Technologies (Peking University), Ministry of Education.
\label{lastpage}

\bibliographystyle{style/ACM-Reference-Format}
\bibliography{paper}

\end{document}